\documentclass{article}
\usepackage{fullpage}
\usepackage{authblk}
\usepackage{amsmath}
\usepackage{amssymb}
\usepackage{hyperref}
\usepackage{graphicx}
\begin{document}

\title{GeoChemFoam: Direct modelling of flow and heat transfer in micro-CT images of porous media}

\author[1]{Julien Maes}

\author[1]{Hannah P. Menke}

\affil[1]{Institute of GeoEnergy Engineering, Heriot-Watt University, Edinburgh, U.K.}

\maketitle

\abstract{GeoChemFoam is an open-source OpenFOAM-based numerical modelling toolbox that includes a range of custom packages to solve complex flow processes including multiphase transport with interface transfer, single-phase flow in multiscale porous media, and reactive transport with mineral dissolution. In this paper, we present GeoChemFoam's novel numerical model for simulation of conjugate heat transfer in micro-CT images of porous media. GeoChemFoam uses the micro-continuum approach to describe the fluid-solid interface using the volume fraction of fluid and solid in each computational cell. The velocity field is solved using Brinkman's equation with permeability calculated using the Kozeny-Carman equation which results in a near-zero permeability in the solid phase. Conjugate heat transfer is then solved with heat convection where the velocity is non-zero, and the thermal conductivity is calculated as the harmonic average of phase conductivity weighted by the phase volume fraction. Our model is validated by comparison with the standard two-medium approach for a simple 2D geometry. We then simulate conjugate heat transfer and calculate heat transfer coefficients for different flow regimes and injected fluid analogous to injection into a geothermal reservoir in a micro-CT image of Bentheimer sandstone and perform a sensitivity analysis in a porous heat exchanger with a random sphere packing.}

\section{Introduction}\label{sec1}

Heat transfer in porous media is of the utmost importance for a range of energy-related applications, including geothermal energy engineering \cite{Grant}, heat exchangers \cite{Xu2017}, nuclear reactors \cite{Wang2010}, in-situ combustion and pyrolysis \cite{Maes2015}, packed-bed reactors \cite{Li2010}, CO$_2$ capture and storage \cite{Aradottir2015}, solar cells \cite{Mohamad2013} and battery technology \cite{2006-Braun}. All of these applications include a range of mechanisms that occur at multiple scales as well as heat transfer at the fluid-solid interface. In addition, the temperature controls processes that may influence the performance of the system, including viscous dissipation \cite{Hooman2008}, chemical reactions \cite{Zhao2000}, phase transfer \cite{Niu2015} and electrical conductivity \cite{Lee2010}. To optimally design for such technologies it is necessary to have a detailed understanding of the transport properties of mass, momentum and energy in porous media at every scale.

Computational Fluid Dynamics (CFD) is an extraordinary tool to design and optimise engineering processes. Applied to 3D X-ray computed micro-tomography images of porous media, flow and transport equations can be solved directly inside the pore-space, i.e. the interconnected void space between the solid grains, offering an unprecedented window into the physics of porous media applications \cite{Alpak2018,Shams2021}. Pore-scale CFD resolves the interface position and concentration gradients exactly, so that the impact of pore-level properties such as surface area, pore-size distribution and contact angle can be investigated and enabling the development of upscaling methods from the micro- to the meso- and the macro-scale. Pore-scale CFD has been successfully applied to investigate hydrodynamic dispersion \cite{Soulaine2021}, multiphase flow \cite{Zhao2019}, single-phase reactive transport \cite{Oliveira2020}, multiphase reactive transport \cite{Maes2021} and electric charge transport \cite{Liu2015}. However, the use of pore-scale CFD to investigate conjugate heat transfer is still in its infancy.

Conjugate heat transfer is traditionally modelled in CFD using a two-medium approach \cite{Quintard1997,Younsi2008}, in which heat transfer in the fluid and solid are solved separately and then coupled by boundary conditions at the fluid-solid interface. This method has been employed successfully to model conjugate heat transfer in open-cell solid foam \cite{Diani2015,Das2016,Yang2021} and micro-CT images of rocks \cite{Song2017}. However this approach has two drawbacks, namely the requirement for an exact correspondence between the fluid and solid boundary cells, and the need to iterate over the full solution scheme to stabilize the coupled boundary conditions between fluid and solid. As a result, all of the studies previously cited only considered a gradient of temperature in the solid during conjugate heat transfer in simplified geometries. In micro-CT images, the solid temperature is usually assumed constant. An alternative solution is to use a single-field approach which solves the flow in the pore-space using a volume-penalizing immersed boundary method \cite{Lopez2012}. Volume penalization implements the no-slip condition on the surface of solid domains through a source term in the momentum equation. Thermal coupling between the solid and fluid phase is solved using a single-field temperature equation, allowing for convective–diffusive transport of heat in the fluid and diffusive transport in the solid. However, the penalization method requires a sharp interface at a very high resolution to model the discontinuities at the fluid-solid boundaries. The micro-continuum approach offers an efficient and robust alternative.

The micro-continuum approach is an extension of the penalization method to model flow directly in unsegmented micro-CT images \cite{Soulaine2016}. This method solves the velocity field with the Brinkman equation, which introduces an additional source term in the momentum equation equal to the permeability of the under-resolved voxel. This permeability is calculated using the Kozeny-Carman equation that depends on a constant related to the image resolution. The micro-continuum approach has been successfully applied to model single-phase flow in a micro-porous rock \cite{Menke2021}, multiphase flow in various porous media geometries \cite{Carrillo2020}, multiphase reactive transport in shales \cite{Soulaine2016b,Soulaine2019}, mineral dissolution in carbonate rocks \cite{Soulaine2017,Soulaine2018,Noiriel2021} and mineral precipitation in simplified pore geometries \cite{Deng2021,Yang2021b}.

The objective of this paper is to extend the micro-continuum approach to model conjugate heat transfer in micro-CT images of porous media. First, the governing equations are presented and our micro-continuum-based approach is described. Next, our model is validated by comparison with the standard two-domain approach for a simple 2D geometry. We then simulate conjugate heat transfer and calculate heat transfer coefficients for different flow regimes and injected fluid in a micro-CT image of Bentheimer sandstone which are analogous to different injection scenarios in a geothermal reservoir. Finally, we perform a sensitivity analysis in a porous heat exchanger with random sphere packing. 

\section{Model description}
\label{sec:1}

\subsection{Governing equations}

The full domain $\Omega$ is made of the fluid domain $\Omega_f$ and the solid domain $\Omega_s$.
The Navier-Stokes equations for the steady laminar motion of an incompressible Newtonian fluid  $\Omega_f$ are \cite{Patankar}
\begin{align}
\nabla\cdot\mathbf{u} & = 0,\label{Eq:cont}\\
\nabla\cdot\left(\mathbf{u}\otimes\mathbf{u}\right) & =-\nabla p +\nu\nabla^2\mathbf{u},\label{Equ:momentum}
\end{align}
with a no-flow, no-slip boundary condition at the fluid-solid interface $\Gamma$,
\begin{align}
&\mathbf{u}\cdot \mathbf{n}_{s}=0 \hspace{0.5cm} \text{at $\Gamma$},\label{Equ:bcu1}\\ 
& \nabla p\cdot \mathbf{n}_{s}=0 \hspace{0.5cm} \text{at $\Gamma$},\label{Equ:bcu2} 
\end{align}
where $\mathbf{u}$ (m/s) is the velocity, $p$ (m$^2$/s$^2$) is the kinematic pressure, $\nu$ (m$^2$/s) is kinematic viscosity and $\mathbf{n_s}$ is the normal vector to the solid surface, pointing toward the fluid. The energy equation in the fluid, written in term of temperature, reads \cite{Diani2015}
\begin{equation}\label{Equ:Tf}
 \nabla\cdot\left(c_fT_f\mathbf{u}\right)=\nabla\cdot\kappa_f\nabla T_f,
\end{equation}
where $c_f$ (kJ/m$^3$/K) is the fluid heat capacity, $T_f$ (K) is the temperature in the fluid and $\kappa_f$ (kW/m/K) is the fluid thermal conductivity. In the solid domain $\Omega_s$, the temperature equation is defined as \cite{SIEGERT2021}
\begin{equation}\label{Equ:Ts}
\nabla\cdot\kappa_s\nabla T_s=0,
\end{equation}
where $T_s$ (K) is the temperature in the solid and $\kappa_s$ (kW/m/K) is the solid thermal conductivity. At the fluid-solid interface, the heat transfer is coupled by a a set of boundary conditions, namely continuity of temperature and continuity of heat flux \cite{Das2016}
\begin{align}
&T_f=T_s \hspace{0.5cm} \text{at $\Gamma$},\label{Equ:bcT1}\\
&\kappa_f\nabla T_f \cdot \mathbf{n}_s= \kappa_s\nabla T_s \cdot \mathbf{n}_s\hspace{0.5cm} \text{at $\Gamma$}.\label{Equ:bcT2}
\end{align}
The two-medium approach solves the velocity field in the fluid phase. Then an iterative scheme is used for the temperature equations, where the fluid temperature (Equ. (\ref{Equ:Tf})) is solved with boundary conditions (Equ. (\ref{Equ:bcT1}) and (\ref{Equ:bcT2})) calculated with a constant temperature in the solid, and the solid temperature (Equ. (\ref{Equ:Ts})) is solved with boundary conditions (Equ. (\ref{Equ:bcT1}) and (\ref{Equ:bcT2})) calculated with a constant temperature in the fluid. This scheme is iterated until convergence \cite{Das2016}.

\subsection{The micro-continuum approach}

The micro-continuum approach uses volume-averaging of the flow equations over a control volume in the presence of solid material. The fluid-solid interface is then described in terms of the volume fractions of fluid $\epsilon_f$ and solid $\epsilon_s$ in each control volume. The heat and mass transport are solved in term of the volume-averaged properties
\cite{Soulaine2016}
\begin{align}
 &\overline{\mathbf{u}}=\frac{1}{V}\int_{V_f}\mathbf{u}dV,\\
 &\overline{p}=\frac{1}{V_f}\int_{V_f}pdV,\\
 &\overline{T}_f=\frac{1}{V_f}\int_{V_f}T_fdV,\\
 &\overline{T}_s=\frac{1}{V_s}\int_{V_s}T_sdV,\\
\end{align}
where $V_f$ and $V_s$ are the volume of fluid and solid in a control volume, and $V=V_f+V_s$.
The volume-averaged velocity satisfies the Darcy-Stokes-Brinkman equation over the full domain $\Omega$ \cite{Soulaine2016}
\begin{align}
\nabla\cdot\overline{\mathbf{u}} &= 0 \label{Eq:contDBS}\\
\frac{1}{\epsilon_f}\nabla\left(\frac{\mathbf{u}}{\epsilon_f}\otimes\mathbf{u}\right)=-&\nabla \overline{p}_f +\frac{\nu}{\epsilon_{f}}\nabla^2\overline{\mathbf{u}}-\nu K^{-1}\overline{\mathbf{u}}\label{Eq:momentumDBS}
\end{align}
where $K$ is the permeability of the control volume. $\nu K^{-1}\overline{\mathbf{u}}$ represents the momentum exchange between the fluid and the solid phase, i.e. the Darcy resistance. This term is dominant in the solid phase and vanishes in the fluid phase. To model this, the local permeability field $K$ is assumed to be a function of the local porosity $\epsilon_f$, following a Kozeny-Carman relationship \cite{Soulaine2016}
\begin{equation}\label{Eq:perm}
 K^{-1}=\frac{180}{h^2}\frac{\left(1-\epsilon_f\right)^2}{\epsilon_f^3},
\end{equation}
where $h$ is the mesh resolution.
The temperature equation is solved in term of the single-field temperature
\begin{equation}\label{Equ:T}
 \overline{T}=\epsilon_f\overline{T}_f+\epsilon_s\overline{T}_s.
\end{equation}
following the transport equation
\begin{equation}
 \nabla\cdot\left(c_f\overline{T}u\right)=\nabla\cdot\overline{\kappa}\nabla\overline{T},
\end{equation}
where $\overline{\kappa}$ is the single-field thermal conductivity, which is calculated as the harmonic average of phase conductivity weighted by the phase volume fraction \cite{Song2017}
\begin{equation}
 \overline{\kappa}=\frac{\kappa_f\kappa_f}{\epsilon_s\kappa_f+\epsilon_f\kappa_s}.
\end{equation}

\section{Upscaling}

The system of equations is only dependent on three dimensionless numbers, namely the Reynolds number, \cite{Patankar}
\begin{equation}
 Re = \frac{UL}{\nu},
\end{equation}
which is the ratio of the inertial to viscous forces, the Prandtl number, \cite{Patankar}
\begin{equation}
 Pr=\frac{c_f\nu}{\kappa_f},
\end{equation}
which is the ratio of momentum to thermal diffusivity, and the conductivity ratio,
\begin{equation}
 R_{\kappa}=\frac{\kappa_s}{\kappa_f}.
\end{equation}
Here $U$ (m/s) and $L$ (m) are the reference velocity and length. For the reference velocity, we use the Darcy velocity
\begin{equation}
 U=\frac{Q}{A},
\end{equation}
where $A$ (m$^2$) is the inlet area and $Q$ (m$^3$/s) is injection flow rate. For the reference length, we use the pore-scale length which, in micro-CT images, can be calculated as
\begin{equation}\label{Eq:length}
 L = \sqrt{\frac{8K}{\phi}},
\end{equation}
where $\phi$ is the porosity and $K$ is the permeability, defined as
\begin{equation}
 K=-\frac{\nu L_DU}{\Delta P},
\end{equation}
where $L_D$ is the full length of the domain and $\Delta P$ the pressure drop between the inlet and outlet. The factor 8 is added so that the pore-scale length corresponds to the tube size for a homogeneous bundle of capillary tubes \cite{PATERSON1983}. Upscaling the heat transfer involves calculating the heat transfer coefficient $k_T$ (kJ/m$^2$/s/K) between the fluid and the solid, defined as \cite{Lopez2012}
\begin{equation}\label{Eq:kT}
 k_T=\frac{\Phi_T}{\Delta TA_s},
\end{equation}
where $\Phi_T$ is the overall conductive heat flux across the interface
\begin{equation}
 \Phi_T=\int_{\Gamma}\kappa_f\nabla T_f\cdot\mathbf{n}_sdS=\int_{\Gamma}\kappa_s\nabla T_s\cdot\mathbf{n}_sdS=\int_{\Omega}\overline{\kappa}\nabla\overline{T}\cdot\nabla\epsilon_f dV,
\end{equation}
$\Delta T$ is the overall temperature difference
\begin{equation}
 \Delta T = \frac{\int_{\Omega_f}T_fdV}{\int_{\Omega_f}dV}-\frac{\int_{\Omega_s}T_sdV}{\int_{\Omega_s}dV}=\frac{\int_{\Omega}\epsilon_f\overline{T}dV}{\int_{\Omega}\epsilon_fdV}-\frac{\int_{\Omega}\epsilon_s\overline{T}dV}{\int_{\Omega}\epsilon_sdV},
\end{equation}
and $A_s$ is the solid surface area \cite{QUINTARD1994}
\begin{equation}
 A_s=\int_{\Gamma}dS=\int_{\Omega}\nabla\epsilon_f\cdot\mathbf{n}_sdV.
\end{equation}
The heat transfer between solid and fluid is then characterized by the Nusselt number, which is the ratio of convective to conductive heat transfer at the fluid-solid interface
\begin{equation}\label{Eq:Nu}
 Nu=\frac{k_TL_D}{\kappa_f}.
\end{equation}

\section{Implementation}
The numerical method is implemented in GeoChemFoam \cite{Maes2021,Maes2020b,Maes2021b}, our reactive transport toolbox based on OpenFOAM\textsuperscript{\textregistered} \cite{OpenFOAM2016}. The full code can be downloaded from \href{www.julienmaes.com}{www.julienmaes.com/geochemfoam/}. The standard solvers \textit{simpleFoam} and \textit{laplacianFoam} were modified into two new solvers \textit{simpleDBSFoam} and \textit{heatTransportSimpleFoam} that respectively solve Equ. (\ref{Eq:momentumDBS}) and Equ. (\ref{Equ:T}) using the SIMPLE algorithm \cite{Patankar}. The equations are discretized on a collocated Eulerian grid. The space discretization of the convection terms is performed using the first-order upwind scheme while the diffusion term is discretized using the Gauss linear limited corrected scheme, which is second order and conservative.

\section{Verification}

\begin{figure}[!b]
\begin{center}
\includegraphics[width=0.5\textwidth]{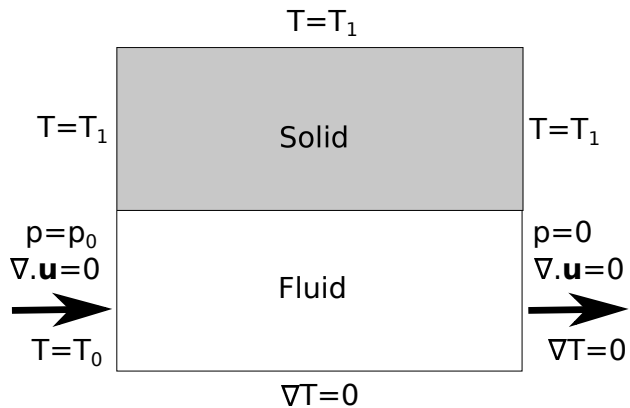}
\caption{Domain and boundary conditions for validation test case\label{fig:2Dgeo}.}
\end{center}
\end{figure}

To validate our method, we compare our numerical results with the standard two-medium approach on a simplified two-dimensional geometry. The domain is a rectangle of size 4 mm $\times$ 5 mm. The bottom half of the domain (0 $<y<$ 2 mm) is a free flow zone and the top half (2 mm $< y <$ 4 mm) is a solid. The geometry and boundary conditions are summarized in Fig. \ref{fig:2Dgeo}. The fluid and solid properties are summarized in Table \ref{Table:param1}.
The steady-state velocity in the free-flow zone has a parabolic profile with a a Darcy velocity that follows Poiseuille equation,
\begin{equation}
 U_D=\frac{L^2}{12\nu}\frac{p_0}{L_D},
\end{equation}
where $L=$ 2 mm is the height of the free-flow zone and $L_D=$ 5 mm is the length of the domain. The inlet pressure $p_0$ is set to $7.5\times10^{-5}$  m$^2$/s$^2$, which gives a Darcy velocity of $5\times10^{-4}$ m/s.  

\begin{table}[!ht]
\centering
\begin{tabular}{c|c|c|c}
Name & Notation & Value & Unit \\[0.1cm]
\hline
& & &\\[-0.2cm]
Fluid viscosity & $\nu$ & $10^{-6}$ & m$^2$/s \\[0.1cm]
Fluid heat capacity & $c_f$ & $4.2\times10^{3}$ & kJ/m$^3$/s \\[0.1cm]
Fluid thermal conductivity & $\kappa_f$ & 0.6 $\times10^{-3}$ & kJ/m/s/K \\[0.1cm]
Solid thermal conductivity & $\kappa_f$ & 6.0 $\times10^{-3}$ & kJ/m/s/K \\[0.1cm]
\end{tabular}
\caption{Fluid and solid properties for validation test case.\label{Table:param1}}
\end{table}

Simulations with both the two-medium approach and the micro-continuum approach are run with increasing mesh resolutions of 200 $\mu$m, 100 $\mu$m, 50 $\mu$m and 25 $\mu$m. For the two-medium approach, the solution procedure is iterated until the maximum change of dimensionless temperature $T^*=T/(T_1-T_0)$ at the fluid-solid interface between the fluid and the solid solutions is smaller than $10^{-5}$.

Table \ref{Table:conv} shows the convergence error for both approaches. The simulations with highest resolution are used for reference and the L$_2$ error is calculated as
\begin{equation}
 L_2=\sqrt{\frac{1}{V_{\Omega}}\int_{\Omega}\left(T^*-T_{ref}^*\right)^2dV}.
\end{equation}
The order two convergence of both methods is clearly visible.

\begin{table}[!ht]
\centering
\begin{tabular}{c|c|c}
Resolution (micron) & $L_2$ error (Two-medium) & $L_2$ error (micro-continuum) \\[0.1cm]
\hline
& & \\[-0.2cm]
200 & 0.07 & 0.15 \\[0.1cm]
100 & 0.02 & 0.04 \\[0.1 cm]
50 &  0.0005 & 0.01 \\[0.1cm]
\end{tabular}
\caption{L$_2$ convergence error with the two-medium approach and the micro-continuum approach for the validation test case.\label{Table:conv}}
\end{table}

\begin{figure}[!t]
\begin{center}
\includegraphics[width=0.9\textwidth]{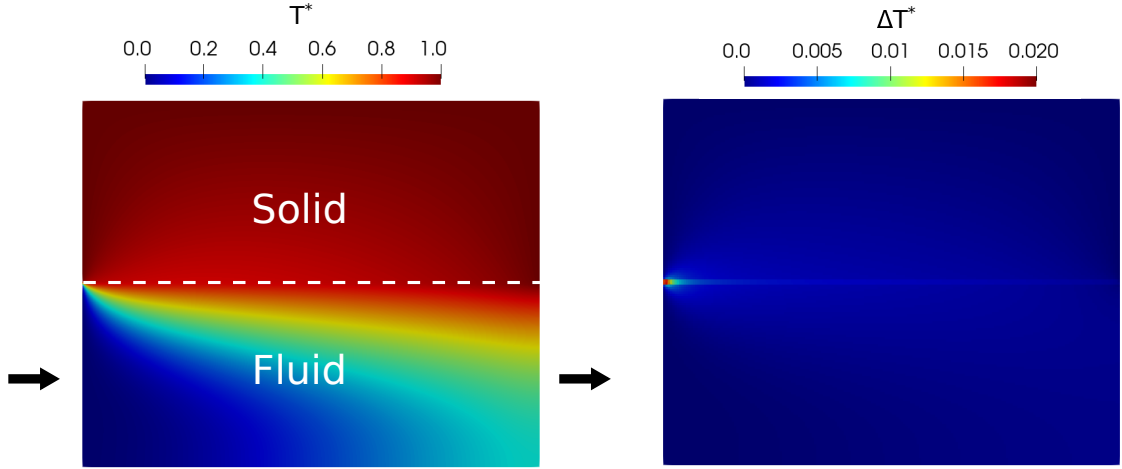}
\caption{Dimensionless temperature map obtained by micro-continuum simulation at 50 $\mu$m resolution (a) and (b) comparison with the results obtained with the two-medium approach during the simulation of conjugate heat transfer in a simplified two-dimensional domain. \label{fig:Tvalid}.}
\end{center}
\end{figure}

Fig \ref{fig:Tvalid}a shows the dimensionless temperature field obtained with the micro-continuum approach at the highest resolution and Fig. \ref{fig:Tvalid}b shows the absolute difference with the one obtained with the two-medium approach. The difference is less than 2\%. 

We conclude that the micro-continuum approach is capable of simulating conjugate heat transfer between solid and liquid with similar accuracy as the two-medium approach and with the same order of convergence (order two).

\section{Applications}

\subsection{Geothermal heat transfer}

Here we consider conjugate heat transfer during injection of cold water into a micro-CT image of Bentheimer sandstone which is analogous to a geothermal reservoir. We perform a series of simulations with various fluid properties and at various Reynolds numbers, and in each case the results are used to calculate the Nusselt number which describe the heat exchange between solid and fluid.

\begin{figure}[!b]
\begin{center}
\includegraphics[width=0.95\textwidth]{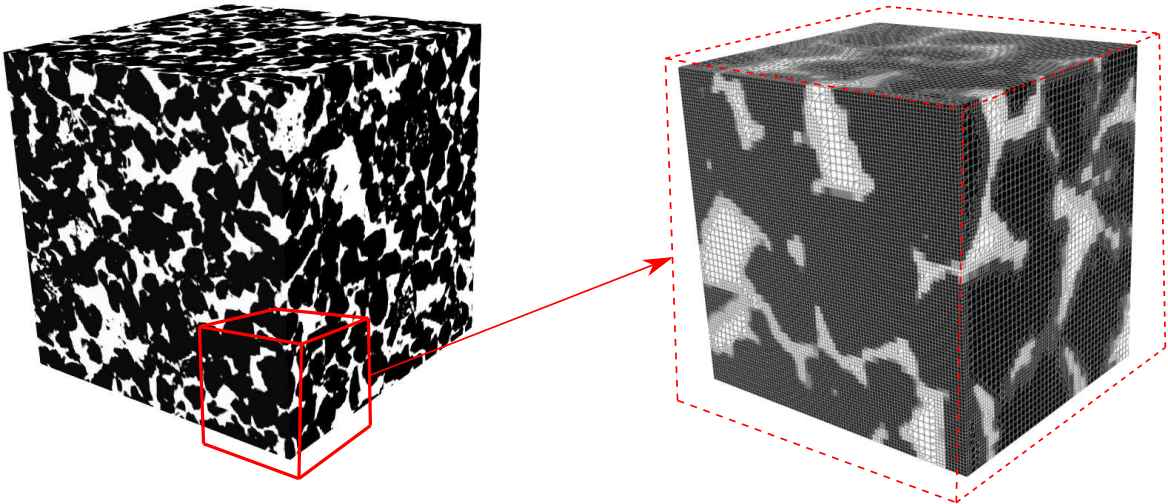}
\caption{Bentheimer micro-CT image and a zoom at the bottom right corner with the two-level mesh used in this study. The solid grains are represented in black, the pores in white and the mesh cells in grey.\label{fig:Bentheimer}}
\end{center}
\end{figure}

The image is a 400$^3$ voxel micro-CT image of Bentheimer sandstone with a resolution of 5 microns. The image is first meshed with a 200$^3$ cells uniform cartesian mesh and for each grid block, $\epsilon_f$ is calculated from the image. Then, each voxel for which $0.01\leq\epsilon_f\leq0.99$ is refined once in each direction and $\epsilon_f$ is recalculated with higher mesh precision. A buffer of length 60 $\mu$m is added on the left to facilitate the injection at constant velocity. We obtain a two-level mesh with 32 million cells and a resolution of 5 $\mu$m around the fluid-solid interface (Fig. \ref{fig:Bentheimer}). The porosity and permeability can be numerically calculated and we obtained $\phi=0.27$ and $K=4.1\times10^{-12}$ m$^2$, which gives a pore-scale length of $1.1\times10^{-5}$ m (Equ. (\ref{Eq:length})).

Fluid is injected from the left boundary at constant velocity and constant temperature $T_0=50$ $^oC$, and exits the domain at the right boundary with constant pressure $p_0$ and no temperature gradient. The top, front, bottom and back boundaries have a no-flow condition and a constant temperature $T_1=70$~$^oC$.

Three different fluids are considered in this study: water H$_2$O, carbon dioxyde CO$_2$ and nitrogen N$_2$. The fluid properties are assumed to be constant, taken at a pressure of 10 MPa and a temperature of 60$^o$ C (\href{https://webbook.nist.gov/chemistry/}{https://webbook.nist.gov/chemistry/}), and summarized in Table \ref{Table:fluidPropBen}. CO$_2$ and N$_2$ are in their supercritical state. The solid thermal conductivity is $3.3\times10^{-3}$ kW/m/K.

\begin{table}[!ht]
\centering
\begin{tabular}{c||c|c|c}
Propertie & H$_2$O & CO$_2$ & N$_2$ \\[0.1cm]
\hline
& & &\\[-0.2cm]
Fluid viscosity (m$^2$)/s & $4.7\times10^{-7}$ & $8.2\times10^{-8}$ & $2.1\times10^{-7}$ \\[0.1cm]
Fluid heat capacity (kJ/m$^3$/K) & $4.1\times10^3$ &$8.8\times10^2$ & $1.1\times10^2$\\[0.1cm]
Fluid thermal conductivity (kW/m/K) & $6.6\times10^{-4}$ & $3.9\times10^{-5}$ & $3.3\times10^{-5}$ \\[0.1cm]
Prandtl number & 2.9 & 1.8 & 0.7 \\[0.1cm]
Conductivity ratio & 5 & 85 & 100 \\
\end{tabular}
\caption{Fluid and solid properties for validation test case.\label{Table:fluidPropBen}}
\end{table}

\begin{figure}[!t]
\begin{center}
\includegraphics[width=0.9\textwidth]{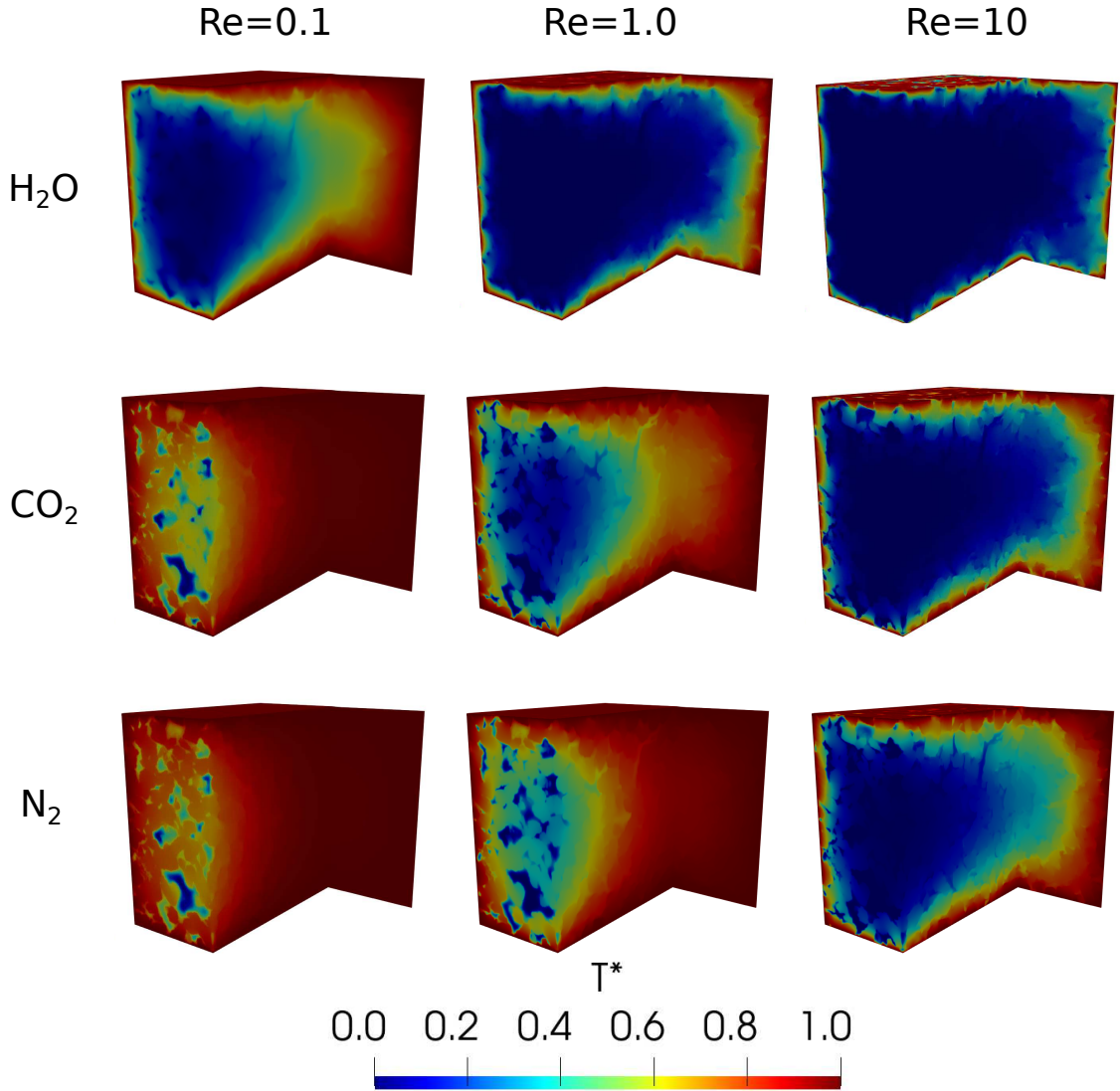}
\caption{Temperature map for different Reynolds numbers and different injected fluids during conjugate heat transfer in a Bentheimer micro-CT image. \label{fig:BentheimerT}}
\end{center}
\end{figure}

For each fluid, the simulation is run at various Reynolds numbers between $Re=0.01$, and $Re=100$ by changing the injection rate. Fig. \ref{fig:BentheimerT} shows the steady-state temperature map of $Re=0.1$, $Re=1.0$ and $Re=10$ for all three injected fluids. The map is shown on a clip from the middle slice y=1 mm and on the last slice x= 2 mm. For all three fluids, the average temperature in the domain decreases as the Reynolds number increases, i.e. as the rate of injection of cold fluid increases. For CO$_2$ and N$_2$, we observe a transition between conduction-dominated heat transfer, for which the injected fluid is quickly heated by the rock, and convection-dominated heat transfer, for which the cold fluid penetrates deeply into the domain and cools down the rock. This transition has already occurred for $Re=0.1$ for H$_2$O, which is characterized by a higher Prandtl number and a much lower conductivity ratio between the solid and the fluid. This transition occurs at a lower Reynolds number for CO$_2$ than for N$_2$, which is characterized by a slightly higher conductivity ratio and a significantly lower Prandtl number.

\begin{figure}[!t]
\begin{center}
\includegraphics[width=0.9\textwidth]{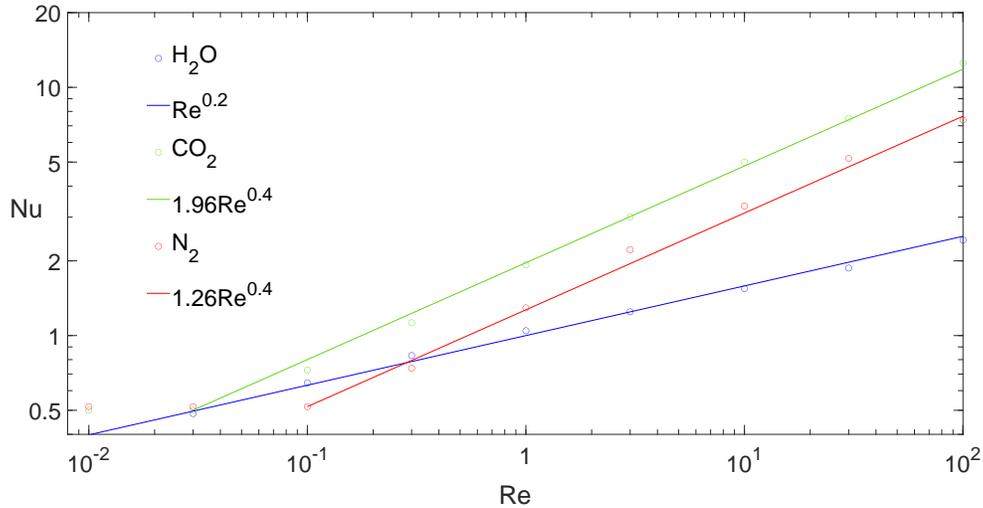}
\caption{The evolution of the Nusselt number as a function of the Reynolds number obtained by numerical simulations and the fitted model during conjugate heat transfer in a Bentheimer micro-CT image for three different fluids. \label{fig:BentheimerNu}}
\end{center}
\end{figure}

In each case, the heat transfer coefficient between the solid and the fluid and the Nusselt number can be calculated (Equ. (\ref{Eq:kT}) and (\ref{Eq:Nu})). Fig. \ref{fig:BentheimerNu} shows the evolution of the Nusselt number obtained by numerical simulations for the three fluids. The figure is plotted on a log-log scale. For H$_2$O, the values obtained fall on a line, which indicates that the Nusselt number is an exponential function of the Reynolds number, and a linear regression gives us $Nu\left(\right.$H$_2$O$\left.\right)=Re^{0.2}$. However, for CO$_2$ and N$_2$, the change of regime between conduction-dominated and convection-dominated transport results in a break with a change of slope where the regime changes. At low Reynolds number, the transport is in the conduction-dominated regime and the Nusselt number is a constant, with $Nu\left(\right.$CO$_2\left.\right)=0.5$ and $Nu\left(\right.$N$_2\left.\right)=0.52$. After the break with a higher Reynolds number, the transport is in the convection-dominated regime, the Nusselt number is an exponential function of the Reynolds number, and  a linear regression gives us $Nu\left(\right.$CO$_2\left.\right)=1.96Re^{0.4}$ and $Nu\left(\right.$N$_2\left.\right)=1.26Re^{0.4}$.

We conclude that our numerical solver is capable of efficiently simulating conjugate heat transfer in a micro-CT image at different regimes and for different fluids. Our numerical results can be used to calculate the Nusselt numbers which describe the heat exchange between solid and fluid, and we obtained the typical results \cite{Lopez2012} with a constant value at low Reynolds numbers in the conduction-dominated regime and an exponential correlation with constant coefficients at high Reynolds numbers in the convection-dominated regime.

\subsection{Porous heat exchanger}

We now consider conjugate heat transfer in a porous heat exchanger. Different injection scenarios, fluid properties and flow regimes are considered and we perform a sensitivity analysis using the design of experiments and Response Surface Methodology (RSM) \cite{Myers} to identify the dominant parameters for the efficiency of the exchanger.

\begin{figure}[!t]
\begin{center}
\includegraphics[width=0.95\textwidth]{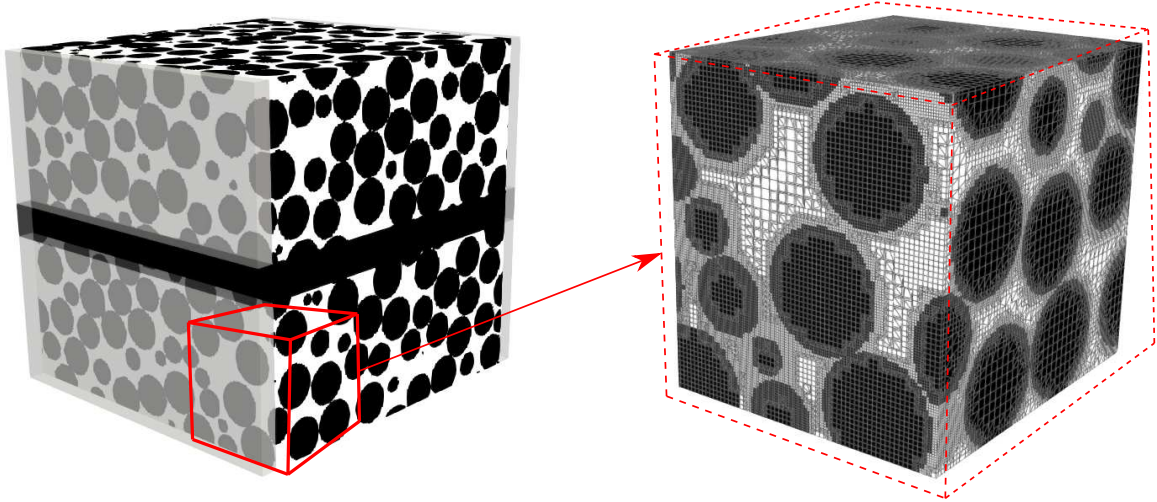}
\caption{Porous heat exchanger micro-CT image and a zoom at the bottom right corner with the three-level mesh used in this study. The solid grains are represented in black, the pores in white and the mesh cells in grey.\label{fig:spherePacking}}
\end{center}
\end{figure}

The domain consists of two random sphere packings of size 20 mm $\times$ 20 mm $\times$ 0.9 mm with identical properties, separated by a solid wall of width $L_w=2$ mm. The spheres are identical with a diameter of 2.5 mm. A buffer of 2 mm on each side of the domain is added to facilitate injection at constant velocity. The domain is given by a micro-CT image of size $440\times400\times400$. The image is first meshed with a uniform cartesian mesh of resolution 0.2 mm and for each grid block, $\epsilon_f$ is calculated from the image. Then each voxel for which $0.01\leq\epsilon_f\leq0.99$ is refined twice in each direction, and their neighbours are refined once, and $\epsilon_f$ is recalculated with higher mesh precision. We thus obtain a three-level mesh with 21 million cells with a resolution of 50 $\mu$m around the fluid-solid interface (Fig. \ref{fig:spherePacking}).  The porosity and permeability is then numerically calculated and we find $\phi=0.366$ and $K=4.75\times10^{-9}$ m$^2$, which gives a pore-scale length of $3.22\times10^{-4}$ m (Equ. (\ref{Eq:length})).

Hot fluid at temperature $T^h_i$ and cold fluid at temperature $T^c_i$ are injected into the bottom and the top part of the domain at constant and equal velocity $U$. $T^h_o$ and $T^c_o$ are defined as the average temperature of the hot and cold fluid at their respective outlets. The performance of the heat exchanger is evaluated in term of its thermal efficiency ratio defined as
\begin{equation}
    \eta = \frac{0.5\left(T^h_i-T^h_o+T^{c}_{o}-T^c_i\right)}{\left(T^{h}_{i}-T^{c}_{i}\right)},
\end{equation}

The sensitivity of the thermal efficiency on selected parameters is investigated. We consider the impact of the Reynolds number $Re$, the Prandtl number $Pr$, the conductivity ratio $R_{\kappa}$ and the flow direction $D$ ($1$ for concurrent  and  $-1$ for countercurrent). A total $n=16$ test cases are simulated with parameters chosen using a full two-level factorial design \cite{Myers}. The simulation parameters and the thermal efficiency ratio calculated from the numerical results are summarized in Table \ref{Table:param}. 

\begin{table}[!ht]
\centering
\begin{tabular}{c||c|c|c|c||c}
Case  & $Re$ & $Pr$ & $R_\kappa$ & $D$ &$\eta$ \\[0.1cm]
\hline
A & 1 & 0.5 & 5 & 1& 0.334 \\
B & 100 & 0.5 & 5 & 1 & 0.0304 \\
C & 1 & 5 & 5 & 1 &   0.115 \\
D & 100 & 5 & 5 & 1 & 0.015\\
E & 1 & 0.5 & 500 & 1& 0.488 \\
F & 100 & 0.5 & 500 & 1 & 0.162\\
G & 1 & 5 & 500 & 1 & 0.356\\
H & 100 & 5 & 500 & 1 & 0.0617 \\
I & 1 & 0.5 & 5 & -1 & 0.383 \\
J & 100 & 0.5 & 5 & -1 & 0.030 \\
K & 1 & 5 & 5 & -1 & 0.130 \\
L & 100 & 5 & 5 & -1 & 0.012\\
M & 1 & 0.5 & 500 & -1 & 0.513 \\
N & 100 & 0.5 & 500 & -1 & 0.162\\
O & 1 & 5 & 500 & -1 & 0.385\\
P & 100 & 5 & 500 & -1 & 0.054 \\[0.1cm]
\end{tabular}
\caption{Simulation parameters and thermal efficiency ratio obtained from numerical simulation for a porous heat exchanger.\label{Table:param}}
\end{table}

The two cases with the highest thermal efficiency are E and M. They correspond to a low Reynolds number, a low Prandtl number and a high conductivity ratio. Concurrently, the two cases with the lowest thermal efficiency are D and L, i.e. the cases with a high Reynolds number, a high Prandtl number and a low conductivity ratio.

\begin{figure}[!t]
\begin{center}
\includegraphics[width=0.9\textwidth]{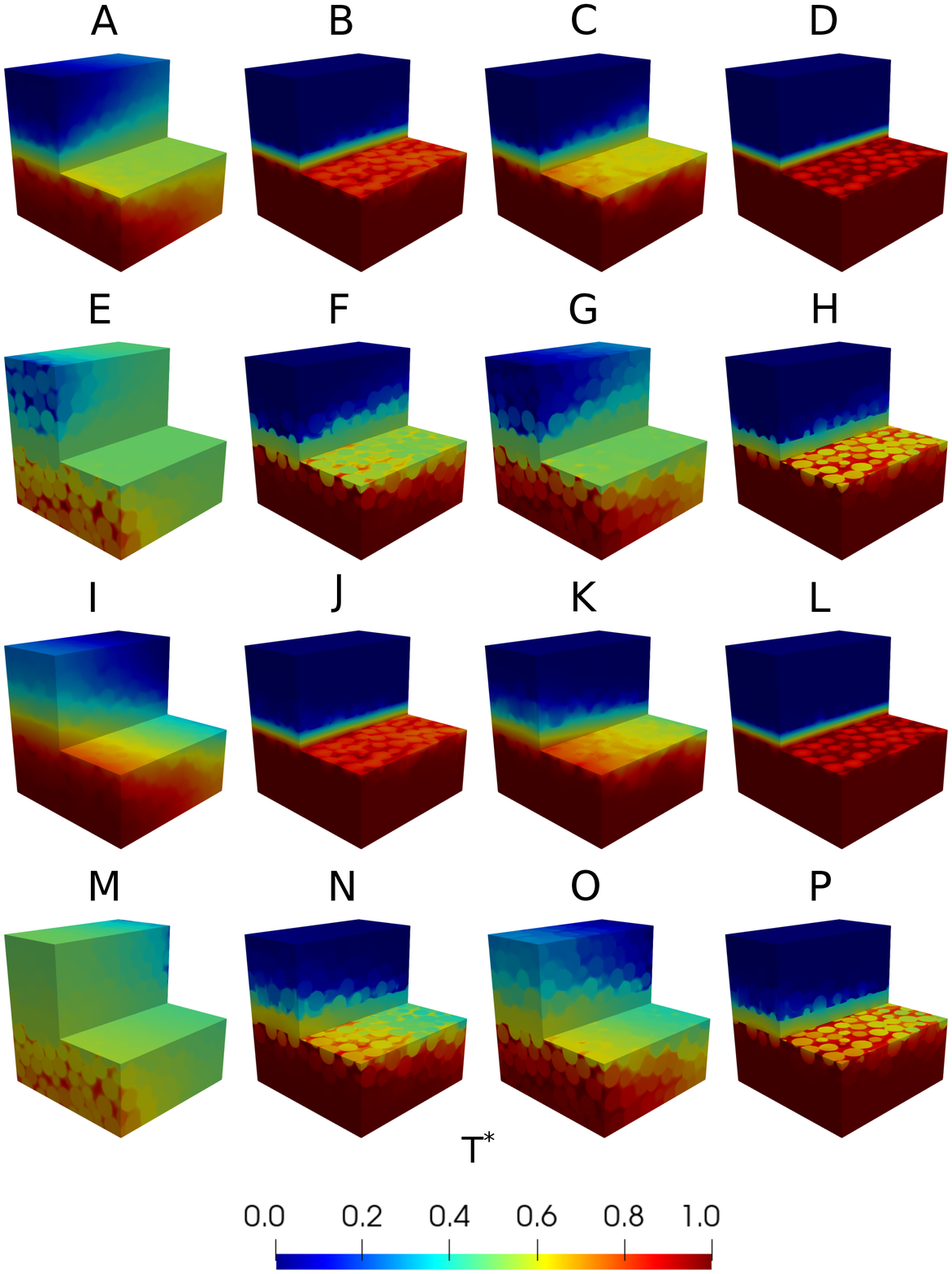}
\caption{The temperature map for all test cases during conjugate heat transfer in a porous heat exchanger\label{fig:spherePackingT}}
\end{center}
\end{figure}

Fig. \ref{fig:spherePackingT} shows the steady-state temperature map for all 16 cases. The map is shown on a clip from the middle slice y=10 mm and on a clip from a slice slightly under the separating wall y=9 mm. We observe that the cases with lower Reynolds number (column 1 and 3), higher conductivity ratio (rows 2 and 3) and lower Prandtl number (column 1 and 2) have a stronger temperature mixing. These cases are in the conduction-dominant regime, either in the fluid, or in the solid, or both. The impact of the flow direction (rows 1 and 2 for concurrent, rows 3 and 4 for counter current) appears to be secondary.

\begin{figure}[!t]
\begin{center}
\includegraphics[width=0.99\textwidth]{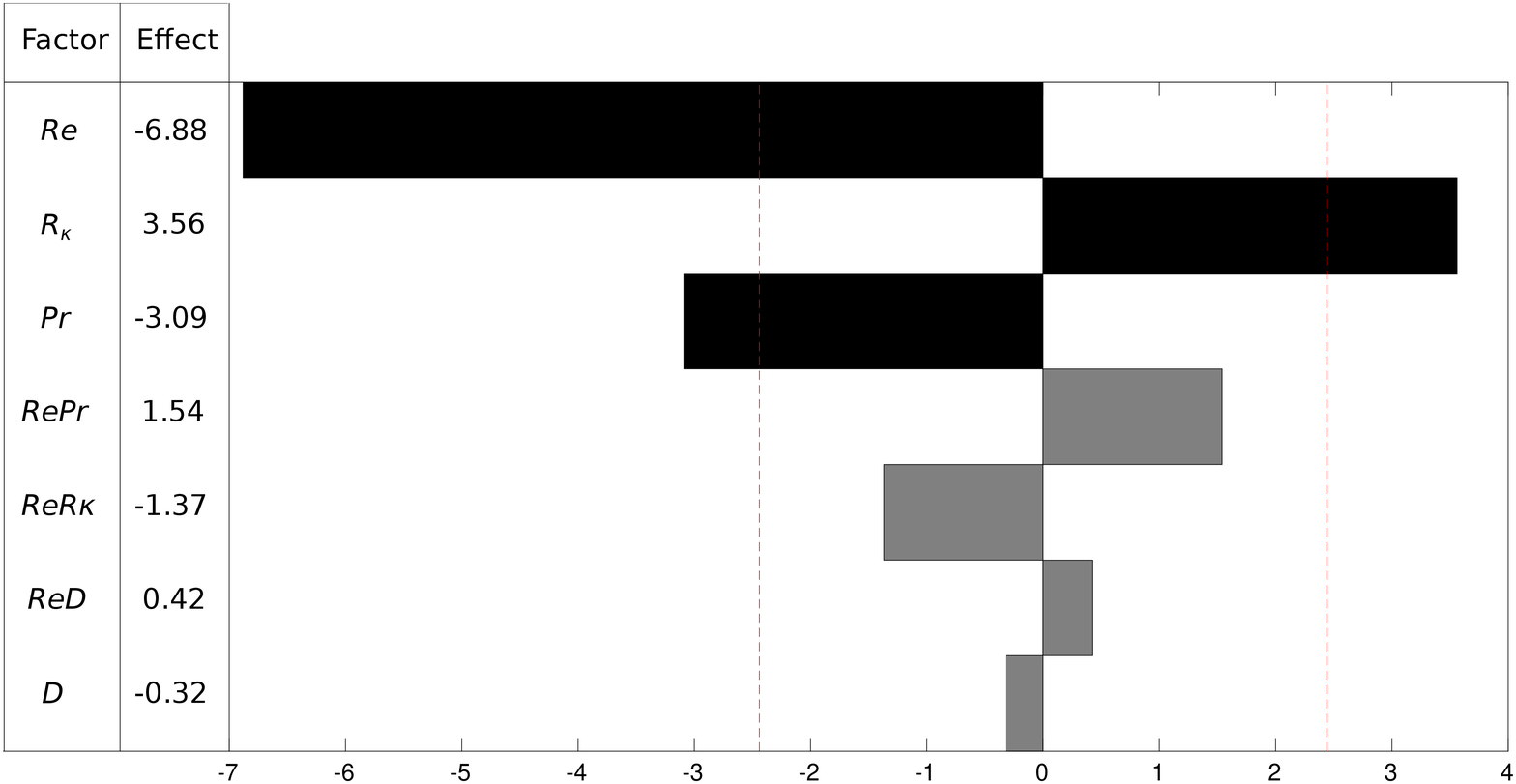}
\caption{A tornado chart of the seven most important factors for the thermal efficiency of a porous heat exchanger. The effects of the primary factors are shown in black, the effects of the secondary factors are shown in grey and the red dashed line show the t-limit for a confidence of 95\%.\label{fig:tornado}}
\end{center}
\end{figure}

Our observation can be confirmed by analysing the parameters impacts using RSM. For this, we consider a first-order response surface model with interaction \cite{Myers}
\begin{equation}
    y_k=\beta_0+\sum\beta_ix_{i,k}+\sum\beta_{ij}x_{i,k}x_{j,k}+\epsilon_k
\end{equation}
where $y_{1\leq k \leq n}$ is the response analysed (here the thermal efficiency ratio), $x_{i,k}$ are the normalized values ($=-1$ or $1$) of the parameters for case $k$ and $\epsilon_k$ the errors.  The $\beta_i$ terms are called the main factor effects and the $\beta_{ij}$ terms the interaction effects. Four main factors and six interactions are considered so a total of $p=10$ effects are calculated. The number of degrees of freedom of the system is defined as $df=n-p=6$. The effects are calculated using a least-square estimator
and characterised in terms of their $t$-value \cite{Myers}
\begin{equation}
    t\left(\beta_i\right)=\frac{\beta_i}{2\sqrt{\frac{MSE}{n}}},
\end{equation}
where the mean squared residual $MSE$ is defined as
\begin{equation}
    MSE=\frac{\sum\epsilon_k^2}{d}.
\end{equation}
The seven most important factors are presented in Fig. \ref{fig:tornado} on a tornado chart and their effects are compared to the $t$-limit value, defined as the 95$^{th}$ percentile of a student $t$-distribution with $df$ degree of freedom
\begin{equation}
    t_{0.05,df=6}=2.45.
\end{equation}

Effects larger than the $t$-limit are called the primary effects and effects lower than the $t$-limit are called the secondary effects. We observe that the system has three primary effects: $Re$ with a negative effect, $R_{\kappa}$ with a positive effect, and $Pr$ with a negative effect. The signs of the effects means that the thermal efficiency decreases as the Reynolds number increases, increases as the conductivity ratio increases, and decreases as the Prandtl number increases, as observed in Table \ref{Table:param} and in Fig. \ref{fig:spherePackingT}. The next two effects are the interaction effects $RePr$ and $ReR_{\kappa}$. The interaction effect $RePr$ is positive while the main effect of $Pr$ is negative. This highlights that the decreases of thermal efficiency with an increase of $Pr$ is more pronounced at low Reynolds number, i.e. in the conduction-dominated regime. Similarly, the interaction effect $ReR_{\kappa}$ is negative while the main effect of $R_{\kappa}$ is positive, showing that the increases of thermal efficiency with the increases of $R_{\kappa}$ is also more significant at low Reynolds numbers. The last two effects are the interaction effect between the Reynolds number and the flow direction and the main effect of the flow direction itself, with similar values and opposite signs. Although their effects are small compared to the three primary parameters, it shows that the thermal efficiency is higher in the countercurrent regime, especially at low Reynolds numbers.

We conclude that our numerical method is suitable for performing sensitivity analysis on a porous heat exchanger, and that our analysis shows that thermal exchange is more efficient in the conduction-dominated regime and for countercurrent flow.

\section{Conclusion}
We present a novel numerical model to simulate conjugate heat transfer in micro-CT images of porous media implemented in GeoChemFoam, our open source reactive transport toolbox. Our model uses a single-field formulation based on the micro-continuum approach to solve the velocity field in the full domain, including the fluid and solid domains, using the Brinkman equation. Conjugate heat transfer is then solved with heat convection where the velocity is non-zero, and the thermal conductivity is calculated as the harmonic average of phase conductivity weighted by the phase volume fractions. 

The model and its implementation were validated by comparing the numerical results for a 2D simplified geometry with the results obtained with a standard two-medium approach. We found that the micro-continuum approach was capable of simulating conjugate heat transfer between solid and liquid with similar accuracy as the two-medium approach and with the same order of convergence (order two).

We then used our numerical toolbox on two test cases. In test case 1, we simulated conjugate heat transfer in a micro-CT image of Bentheimer sandstone with injection scenarios relevant to geothermal reservoirs. Simulations were performed at various Reynolds numbers in order to investigate the evolution of the Nusselt numbers which describe the heat exchange between solid and fluid. We obtained the typical results \cite{Lopez2012} with a constant value at low Reynolds numbers in the conduction-dominated regime and an exponential correlation with constant coefficients at high Reynolds numbers in the convection-dominated regime.

Finally, in test case 2 we simulated conjugate porous heat transfer with a random sphere packing. We performed a sensitivity analysis using RSM considering four parameters: the Reynolds number, the Prandtl number, the conductivity ratio and the flow direction. A total of 16 cases were simulated and the effect of each of the parameters and their interactions were evaluated. Our analysis showed that the Reynolds number, the conductivity ratio and the Prandtl number were the dominant parameters in this order.

The numerical methods and toolbox presented in this work are readily applicable to several engineering applications such as heat transfer in a geothermal reservoirs and porous heat exchangers. However, it is the extension of this method to more complex physics where several processes are combined that will deliver the highest impact. The micro-continuum approach has been applied within the scientific community to many processes including multiphase flow \cite{Carrillo2020}, multiphase reactive transport \cite{Soulaine2019}, mineral dissolution in carbonate rocks \cite{Noiriel2021} and mineral precipitation in simplified pore geometries \cite{Yang2021b}. All of these methods have already been developed within GeoChemFoam \cite{Maes2020b,Maes2021} or are currently under-development, thus enabling the combination of all of these processes within a single model and extending the capabilities of CFD within the scientific community. Furthermore, the micro-continuum approach can also be employed to simulate reactive transport in porous media at the Darcy-scale \cite{Golfier2002}, paving the way for multiscale modelling of coupled thermal and reactive transport processes such as enhanced geothermal systems at the large scale that includes information at every relevant scale integrated using either effective parameter correlations \cite{Lichtner2007} or machine-learning \cite{Menke2021}.

\section*{Supplementary information}

The micro-CT images used in this study are provided in supplementary materials.

\section*{Declarations}

\begin{itemize}
\item Conflicts of interest/Competing interests
The authors declare no competing interests

\item Availability of data, code and material
All data, code and material are available online at www.julienmaes.com/geochemfoam
\end{itemize}

\bibliographystyle{spphys}
\bibliography{References}% common bib file


\end{document}